\documentclass[a4paper,serif,12pt]{article}
\usepackage{geometry}
\usepackage{makeidx}
\usepackage{graphicx}
\usepackage[dvipsnames,svgnames,table]{xcolor}
\usepackage{epstopdf}
\usepackage{epsfig}
\usepackage[font=small]{caption}
\usepackage{textcomp}
\usepackage{hyperref}
\usepackage{amsmath}
\usepackage{amssymb}
\usepackage{multirow}
\usepackage{multicol}
\usepackage{booktabs}
\usepackage{lastpage}
\usepackage{hyperref} 
\usepackage{enumerate}
\usepackage{tabularx}
\usepackage{threeparttable}
\usepackage{float}
\usepackage{array}
\usepackage{cite}
\makeatletter
\renewcommand\section{\@startsection{section}{1}{\z@}%
                                       {-2.5ex \@plus -1ex \@minus -.2ex}%
                                       {2.3ex \@plus.2ex}%
                                       {\normalfont\large\bfseries}}
\renewcommand\subsection{\@startsection{subsection}{1}{\z@}%
                                       {-2.5ex \@plus -1ex \@minus -.2ex}%
                                       {2.3ex \@plus.2ex}%
                                       {\small\bfseries}}
\makeatother
\begin{document}
\begin{center}
\large{\textbf{3D Surface Radiation Dosimetry of a Nuclear-Chicago NH3 Neutron Howitzer}}
\end{center}
\bigskip
\begin{center}
\noindent\textbf{Ahmet Ilker Topuz, Iskender Atilla Reyhancan }\\
\medskip
Istanbul Technical University, Energy Institute, 34469 Istanbul, Turkey\\
E-mail address:  topuz15@itu.edu.tr, aitopuz@gmail.com\\
Tel:  +90 534 020 70 00\\
\begin{abstract}
The tracking of radiation dose is a fundamental concept for the safe operation of nuclear instruments, and this concept is a multi-task challenge in the case of neutron sources since not only the neutron emissions from the main radioactive source but also the neutron capture gamma-rays as well as the gamma-rays from the inelastic neutron scattering make contribution to the total dose received by the facility personnel. The present study is devoted to the assessment of the equivalent dose rates on the surface of a Nuclear-Chicago NH3 neutron howitzer with a 5-Ci ${}^{239}$Pu-Be neutron source by means of Monte Carlo simulations. The 5-Ci ${}^{239}$Pu-Be neutron source is mounted within a vertical aluminum channel at the center of the neutron howitzer, and we use the axial symmetry of the aluminum drum filled with an ordinary paraffin without any potent neutron absorber such as boron or cadmium. We utilise the ICRP-21 flux-to-dose rate conversion factors and we track the neutron and photon equivalent dose rates on the lateral and top surfaces including the outlets of the vertical and horizontal irradiation channels by the aid of point detectors in the MCNP5 code. From the results over a number of sub-regions on the semi-cylindrical drum of the Nuclear-Chicago NH3 neutron howitzer, we show that the outlets of both the vertical and horizontal irradiation channels deliver approximately the same level of equivalent dose rates, and the maximum equivalent dose rates for neutrons and photons on the lateral surface are obtained at the spatial coordinates aligned with the location of the 5-Ci ${}^{239}$Pu-Be neutron source, whereas the periphery around the source channel on the top surface of the neutron howitzer constitutes the highest level of equivalent dose rates among the tracked locations.
\end{abstract}
\end{center}
\section{Introduction}
Side effects of exposure to ionizing radiation are the primary concerns about the safety of occupational personnel during the course of operations with the radioactive sources. Depending on the nature of the ionizing radiation, the quantitative information for the weighted amount of deposited energy in the human body is given by the equivalent dose, and this quantity is monitored for the facility staff  in order to maintain the feasibility of operations under the permitted limits as well as implement the countermeasures in the case of emergency. In principal, the dosimetric standards set by the authorities are to be fulfilled when coming in close contact with the nuclear instruments~\cite{radioprot}.

Neutron howitzer, which is a product from the initial decades of the atomic age containing the isotopic neutron sources~\cite{PuBeStr1,PuBeStr2,isotopic}, is among the nuclear instruments utilised for the chemical analysis~\cite{Howit1,Howit2,Howit3}. Nuclear-Chicago NH3 also belongs to the family of neutron howitzers, and generally, a ${}^{239}$Pu-Be source of certain activity is used for the emission of neutrons by means of ($\alpha$,n) reaction in a cylindrical drum~\cite{chic1,chic2,chic3}. The shielding concept followed in the NH3 howitzer is the thermalisation of neutrons through the utilisation of ordinary paraffin as a moderating medium. However, this strategy results in the generation of photons, and the total equivalent dose received by the occupational staff becomes a function of both neutrons and photons leaking out the NH3 howitzer.   

In the present study, we perform Monte Carlo simulations (MCS) by using the MCNP5 code to calculate the equivalent dose rates (EDRs) for neutrons as well as photons on the cylindrical drum of our Nuclear-Chicago NH3 howitzer with a 5-Ci ${}^{239}$Pu-Be neutron source~\cite{mcnp}. First, we define a 3D geometry by taking into account the dimensional features of the NH3 components. Then, we distribute a number of point detectors~\cite{briesmeister} over the semi-cylindrical drum by taking advantage of the axial symmetry, whereupon we use the ICRP-21 flux-to-dose rate conversion factors in order to obtain the values of the EDRs on the surface of the NH3 howitzer~\cite{ICRP21}. Finally, we deliver the map of the determined EDRs including the critical regions such as the outlets of both the vertical and horizontal irradiation channels over the contact surface of the NH3 howitzer.
\section{Nuclear-Chicago NH3 howitzer with 5-Ci ${}^{239}$Pu-Be}
\begin{figure}[H]
\begin{center}
\setlength{\belowcaptionskip}{-4ex} 
\includegraphics[width=7.4cm]{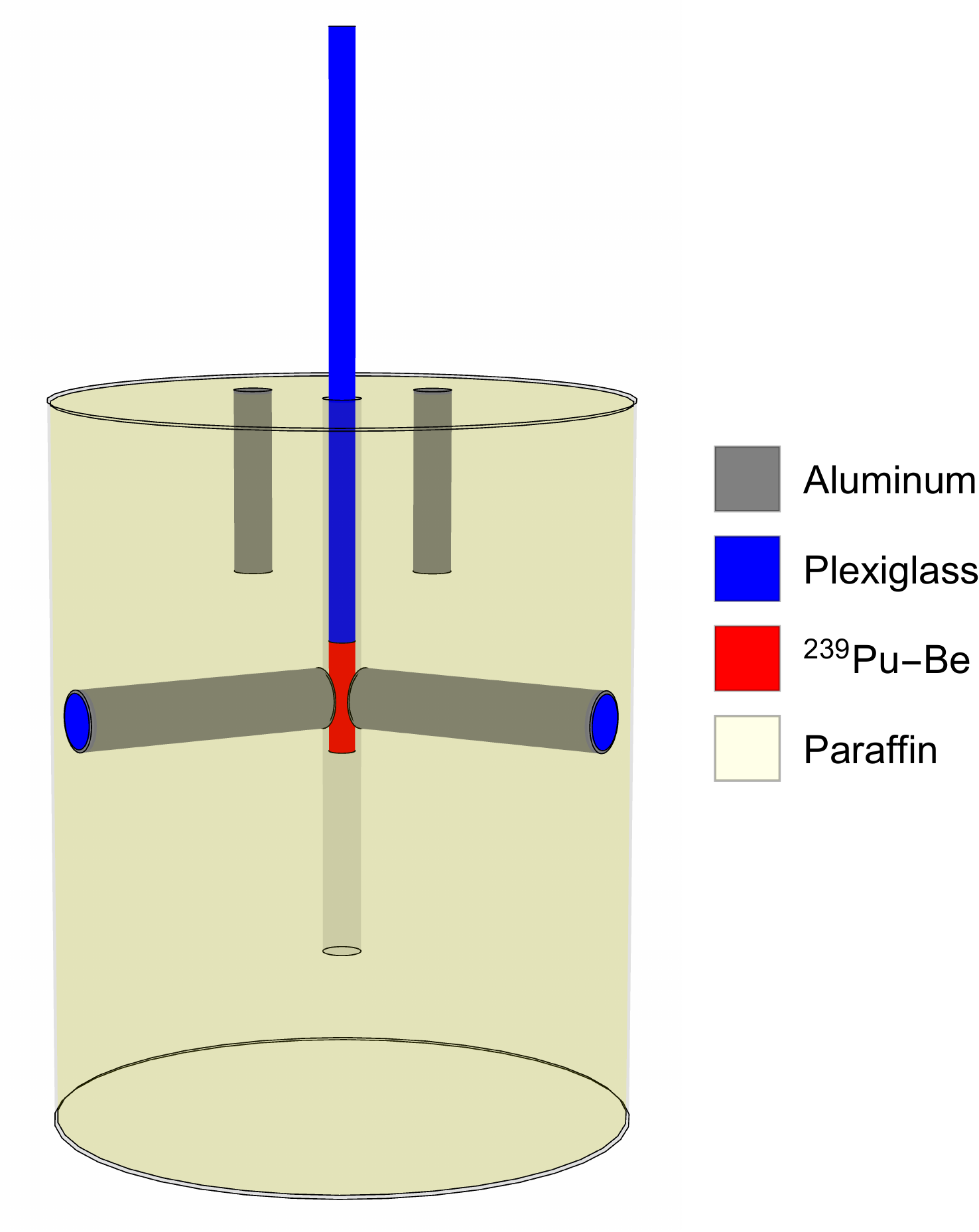}
\caption{Scheme of the Nuclear-Chicago NH3 howitzer with the 5-Ci ${}^{239}$Pu-Be source.}
\label{Howitzer}
\end{center}
\end{figure}

\begin{figure}[H]
\begin{center}
\setlength{\belowcaptionskip}{-4ex} 
\includegraphics[width=13cm]{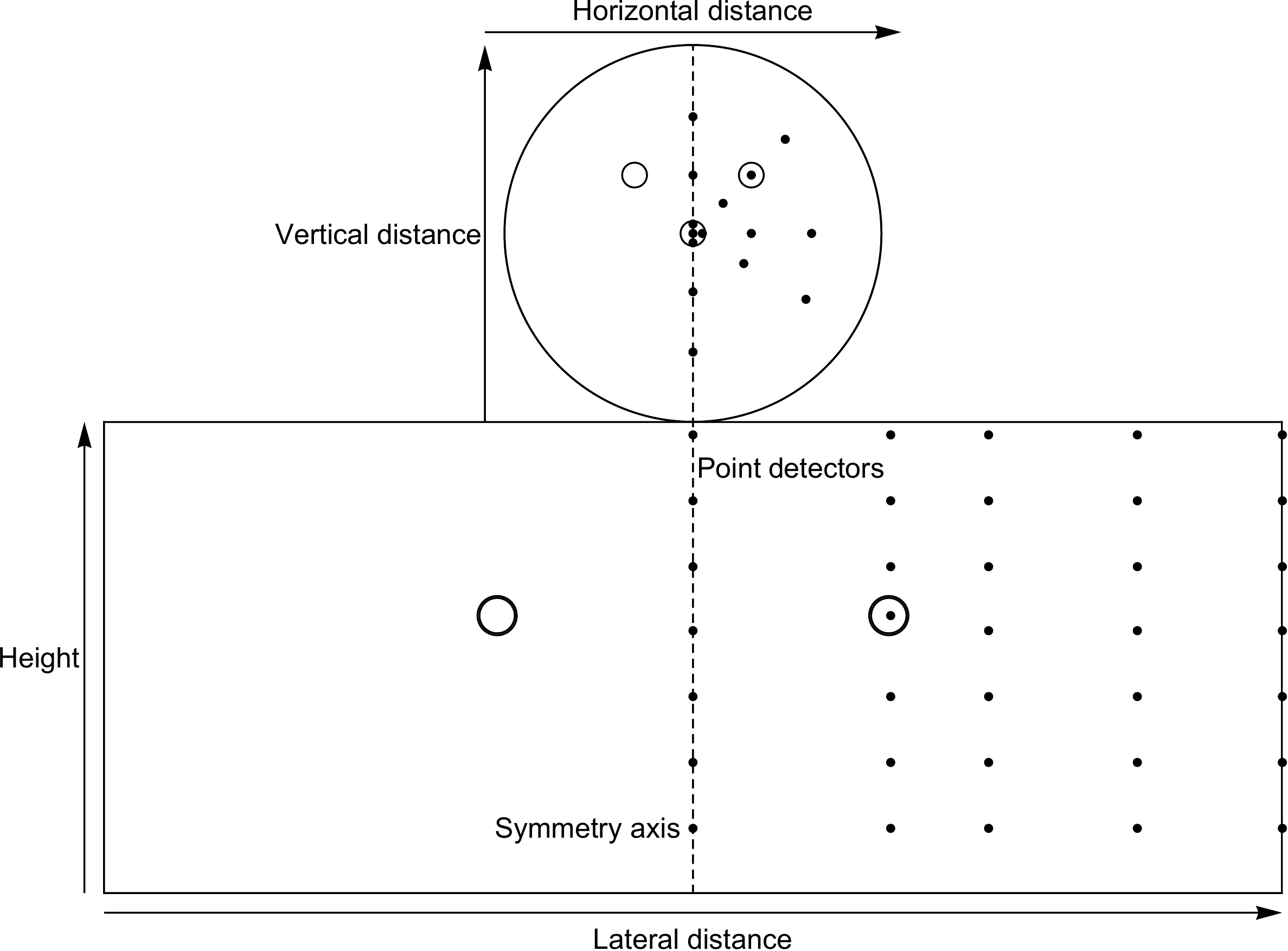}
\caption{Net of the NH3 howitzer consisting of the top surface and the lateral surface.}
\label{decomp}
\end{center}
\end{figure}

The Nuclear-Chicago NH3 howitzer is a cylindrical drum made of aluminum that consists of two horizontal irradiation channels, two vertical irradiation channels, and a vertical source channel. The horizontal irradiation channels are separated by an angle of 120$^\circ$, and the source channel in combination with the two vertical irradiation channels constitutes an isosceles right triangle. Ordinary paraffin (15.7 wt.$\%$H+84.3 wt.$\%$C) infills the cylindrical drum and it functions as the moderating medium for the neutron thermalisation. The vertical irradiation channels are left empty, whereas plexiglass (8 wt.$\%$H+32 wt.$\%$O+60 wt.$\%$C) bars are plugged into the horizontal irradiation channels. The source channel contains the 5-Ci ${}^{239}$Pu-Be that is hold by a plexiglass rod. The scheme of the NH3 howitzer is illustrated in Fig.~\ref{Howitzer}. 
The geometrical and dimensional properties of the Nuclear-Chicago NH3 components are listed in Table~\ref{geodim}. The material card in our MCNP input file is created according to the corresponding values in the compendium of material composition data from the PNNL~\cite{compendium}.
\begin{table} [H]
\begin{footnotesize}
\begin{center}
\begin{threeparttable}
\captionsetup{skip=0pt}
\caption{Geometrical and dimensional properties of the Nuclear-Chicago NH3 components.}
\begin{tabular}{*5c}
\toprule
Components&Material&Geometry&Radius (cm)$\times$Height (cm)&Thickness (cm)\\
\midrule
Drum& Aluminum&Closed cylinder&28.8\tnote{a} $\times$72 &0.3\\
Moderator& Paraffin&Closed cylinder&28.5$\times$71.4 &57\tnote{b} \\
Source channel&Aluminum&Open cylinder&1.9\tnote{a} $\times$55.3 &0.3\\
Source rod&Plexiglass&Closed cylinder&1.3$\times$57.8&2.6\tnote{b}\\
Vertical channel&Aluminum&Open cylinder&1.9\tnote{a} $\times$18.3&0.3\\
Horizontal channel&Aluminum&Open cylinder &3\tnote{a} $\times$27.2&0.3\\
Horizontal channel bar&Plexiglass&Closed cylinder& 2.7$\times$26.9&5.4\tnote{b}\\
\bottomrule
\end{tabular}
\label{geodim}
\nointerlineskip
\begin{tablenotes}
\item[a] Outer radius
\item[b] Diameter
\end{tablenotes}
\end{threeparttable}
\end{center}
\end{footnotesize}
\end{table}
\vspace*{-\baselineskip}
In the present study, the center of the 5-Ci ${}^{239}$Pu-Be neutron source is made aligned with the horizontal channel, which corresponds to a height of 42.5 cm from the bottom of the Nuclear-Chicago NH3 howitzer. The dimension of the neutron source is 1.02 in.$\times$4.425 in.~\cite{chic2}. The neutron strength, which is used for the normalisation of flux values, is equal to 1.6$\times10^{6}$   $\rm s^{-1}Ci^{-1}$~\cite{PuBeStr1,PuBeStr2}. The source is biased by using the radial and axial sampling weights~\cite{Shultis}. The energy spectrum lies in the interval between 0 and 10.5 MeV with the mean energy ($E_{\rm mean}$) equal to 4.24 MeV~\cite{PuBespec}. The source properties are summarised in Table~\ref{sourceprop}.  
\begin{table} [H]
\begin{footnotesize}
\begin{center}
\begin{threeparttable}
\captionsetup{skip=0pt}
\caption{Properties of the 5-Ci ${}^{239}$Pu-Be in the present study.}
\begin{tabular*}{\columnwidth}{@{\extracolsep{\fill}}*5c}
\toprule
Source&Geometry&Activity (Ci)&Strength ($\rm s^{-1}Ci^{-1}$)&Diameter (in.)$\times$Height (in.)\\
\midrule
${}^{239}$Pu-Be&Closed cylinder&5&1.6$\times10^{6}$&1.02$\times$4.425\\
\midrule
\end{tabular*}
\begin{tabular*}{\columnwidth}{@{\extracolsep{\fill}}*5c}
\midrule
Radial weighting&Axial weighting&Spectrum (MeV)&Major peaks (MeV)&$E_{\rm mean}$ (MeV)\\
\midrule
1\tnote{a} & 0\tnote{a} & [0,10.5] & 3.1,4.5& 4.24\\
\bottomrule
\end{tabular*}
\label{sourceprop}
\nointerlineskip
\begin{tablenotes}
\item[a] Power law
\end{tablenotes}
\end{threeparttable}
\end{center}
\end{footnotesize}
\end{table}
\vspace*{-\baselineskip}
For the sake of simplicity, we use a 2D geometrical representation for the calculated EDRs on the 3D surface of the NH3 howitzer as shown in Fig.~\ref{decomp}. This representation simply is the net of the cylindrical drum excluding the bottom circular surface. Thus, our representation is founded on two fundamental contact surfaces: the top surface and the lateral surface. The top surface is a circular surface that includes the outlets of all the vertical channels and also the cross-sectional area of the plexiglass rod in the source channel. The spatial coordinates on the top surface are labeled as vertical distance and horizontal distance. The interval of both these coordinates is [-radius,+radius], which is apparently equivalent to [-28.8 cm, 28.8 cm]. The lateral surface is a rectangular surface that accommodates the outlets of the horizontal channels and the cross-sectional area of the plexiglass bars in these channels. The spatial coordinates on the lateral surface are tagged as height and lateral distance. While the height varies between 0 and 72 cm, the interval of the lateral distance is [-90.5 cm, 90.5 cm] since its length equals $\rm 2\pi\times radius$.
The presence of the symmetry axis in Fig.~\ref{decomp} assists in the reduction of the computation cost, hence we distribute the point detectors~\cite{briesmeister}, which are also called next-event estimators, on only one half of the cylindrical drum. The radius of these point detectors is set as 0.2 cm~\cite{Shultis}. The photon model in the present study, which adopts the default settings in the MCNP5 code, excludes the photoneutron production. We ignore the contribution from the photons when the energies fall below 10 keV. The number of histories utilised in all the simulation cases is $5\times 10^{6}$. Under this setup, by using the ICRP-21 flux-to-dose rate conversion factors, the flux values obtained from the point detectors are converted to the equivalent dose rates expressed in \textmu Sv/h~\cite{ICRP21}.
\section{Profiles of equivalent dose rates}
\subsection{The lateral surface}
\begin{figure}[H]
\begin{center}
\setlength{\belowcaptionskip}{-4ex} 
\includegraphics[width=12cm]{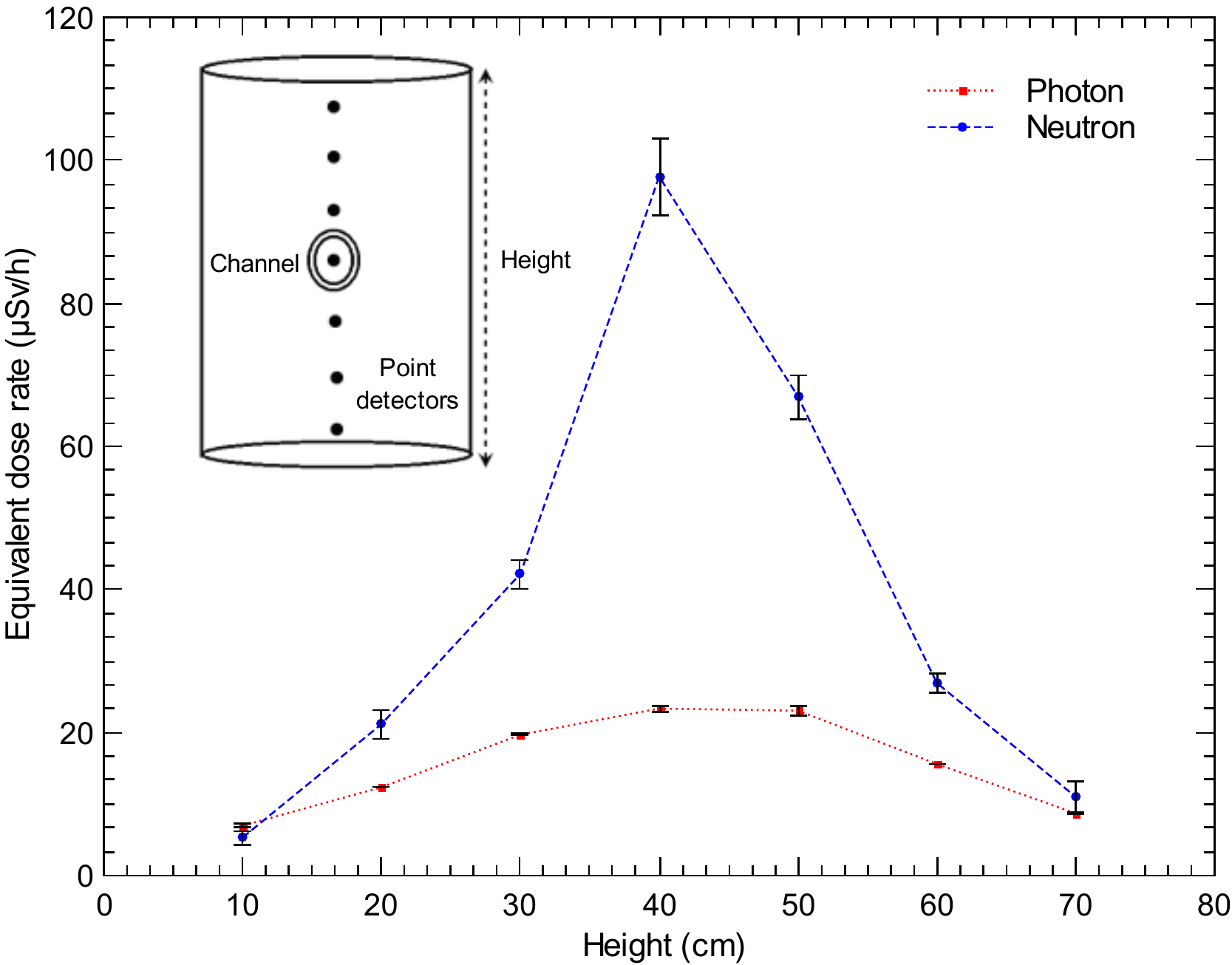}
\caption{Axial variation of EDR along the outlet of the horizontal channel by using seven point detectors.}
\label{bottomplot}
\end{center}
\end{figure}

\begin{figure}[H]
\begin{center}
\setlength{\belowcaptionskip}{-4ex} 
\includegraphics[width=10.5cm]{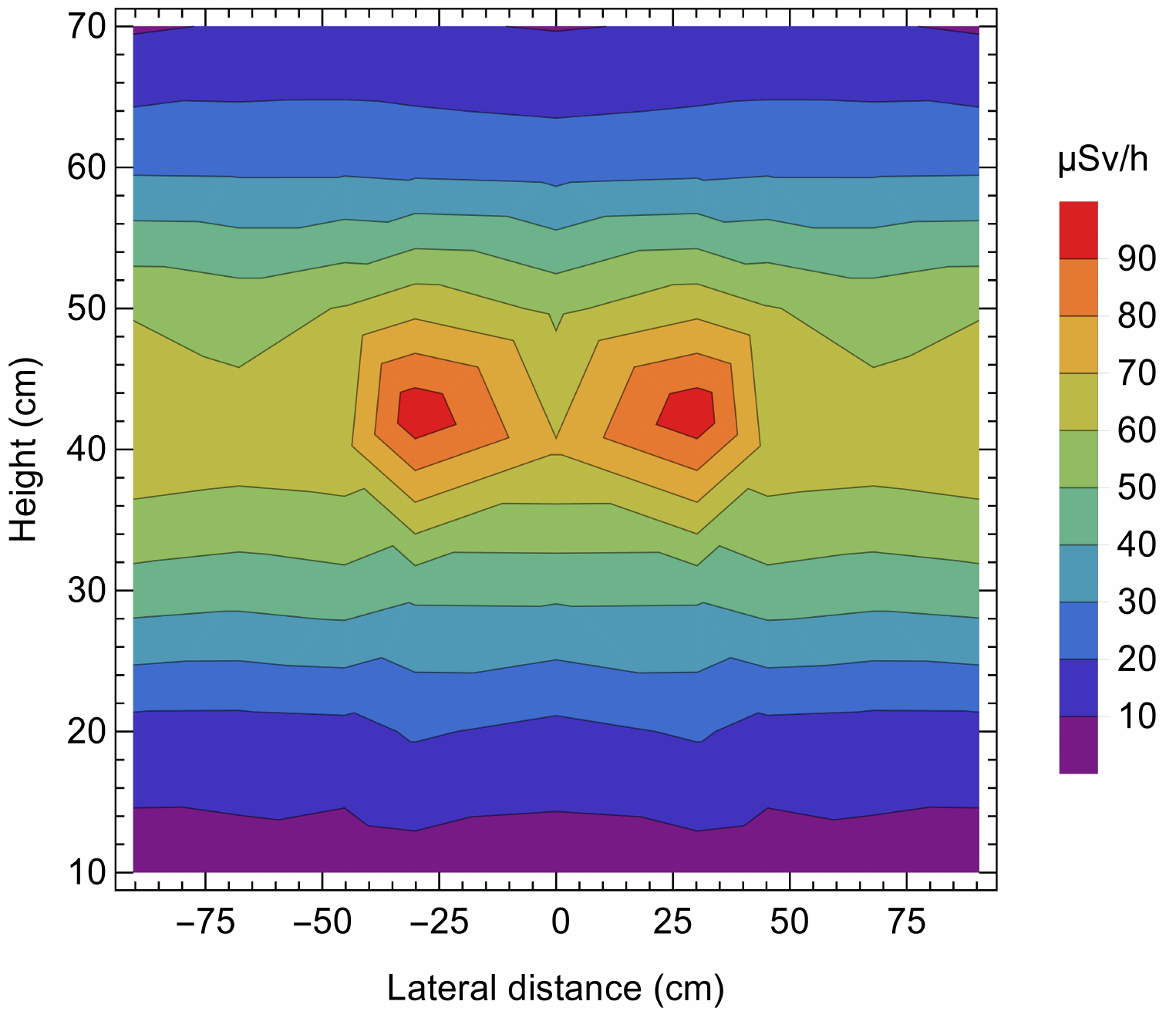}
\caption{Contours of EDR for neutrons on the lateral surface of the NH3 howitzer.}
\label{doselateralneut}
\end{center}
\end{figure}

\begin{figure}[H]
\begin{center}
\setlength{\belowcaptionskip}{-4ex} 
\includegraphics[width=10.5cm]{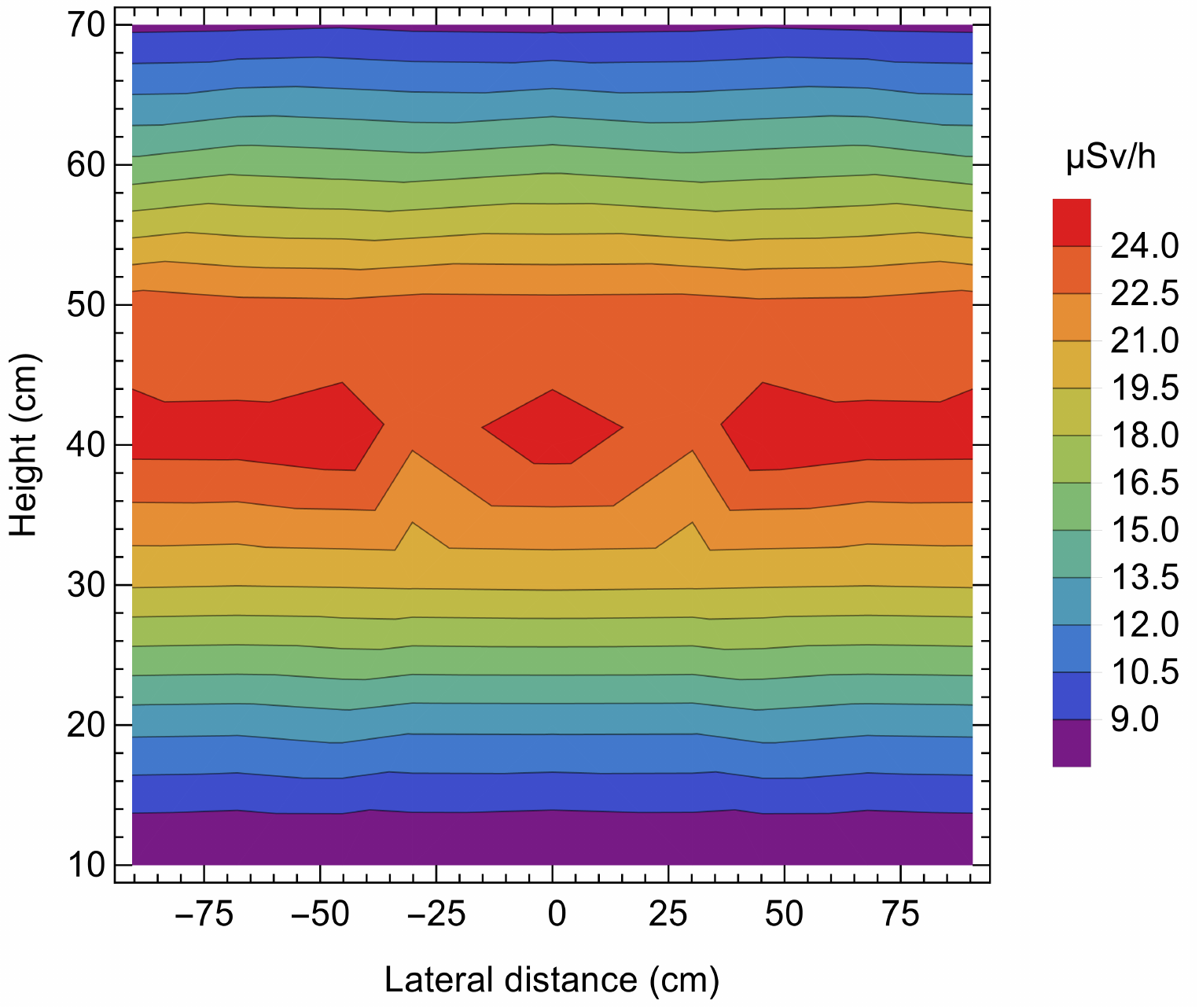}
\caption{Contours of EDR for photons on the lateral surface of the NH3 howitzer.}
\label{doselateralphot}
\end{center}
\end{figure}

\begin{figure}[H]
\begin{center}
\setlength{\belowcaptionskip}{-4ex} 
\includegraphics[width=10.5cm]{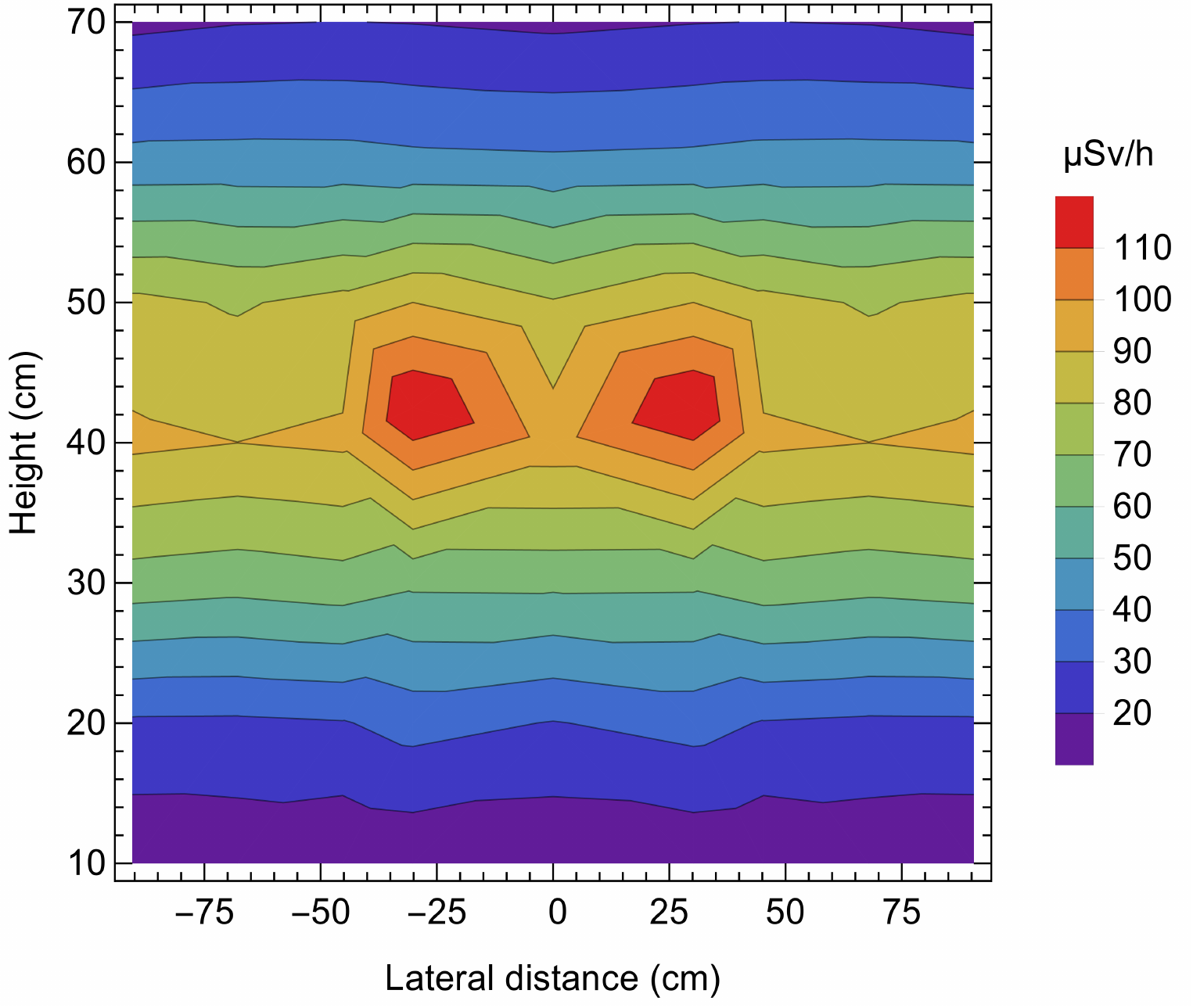}
\caption{Contours of total EDR on the lateral surface of the NH3 howitzer.}
\label{doselateraltotal}
\end{center}
\end{figure}

\begin{figure}[H]
\begin{center}
\setlength{\belowcaptionskip}{-4ex} 
\includegraphics[width=10.5cm]{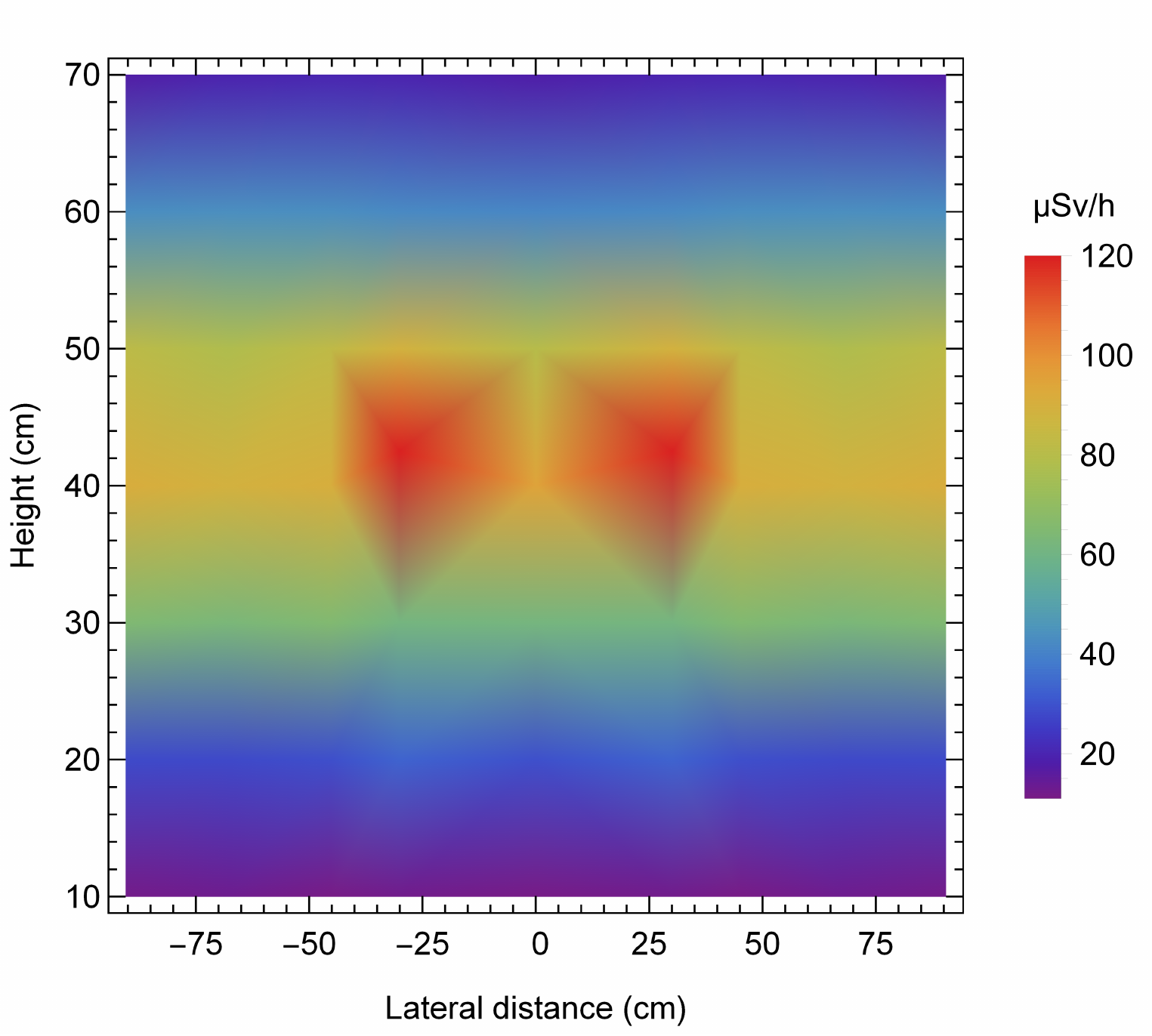}
\caption{Density of total EDR on the lateral surface of the NH3 howitzer.}
\label{dosedensitylateraltotal}
\end{center}
\end{figure}

One of the most critical regions, which is frequently used during the operation with the NH3 howitzer, is the outlet of the horizontal irradiation channel and its periphery. Thus, we initially check the axial variation of EDR along the exit of the horizontal channel through the MCS. Fig.~\ref{bottomplot} shows the EDR values determined by seven point detectors for neutrons and photons. While the EDRs for both neutrons and photons show a parabolic trend, the EDR of neutrons mainly governs the resultant EDR around the circumference of the horizontal channel, reaching a value which is almost four times higher than that of photons at the edge of the plexiglass bar. More specifically, the EDR for neutrons at the outlet of the horizontal channel is 97.735$\pm$5.375 \textmu Sv/h, whereas it is 23.344$\pm$0.443 \textmu Sv/h for photons. 
A more delicate way for demonstrating the EDR values is to illustrate the contour plots over the surfaces defined in Fig.~\ref{decomp}. First, we show the contours of EDR for neutrons over the lateral surface of the NH3 howitzer in Fig.~\ref{doselateralneut}. The EDR patterns for neutrons indicate that the dose accumulation takes place around the periphery of the horizontal channels in elliptical forms, and the EDR for neutrons decreases when moving away from the outlets of channels on the lateral surface. 
Secondly, we depict the EDRs for photons in Fig.~\ref{doselateralphot} and we prefer a finer partitioning for the contour scale to unveil the outlets of the horizontal channels; otherwise, the EDRs of photons exhibit clear bands for a coarser division. The arrow-like patterns point out the locations of the horizontal channels, and it is clearly seen that a region of high intensity extends along the lateral distance at the level of the horizontal channels.  
Finally, we superpose the EDRs of photons onto the EDRs of neutrons as shown in Fig.~\ref{doselateraltotal} and we reveal that the resultant EDRs appear to be similar to those of neutrons. Reminding that the center of the 5-Ci ${}^{239}$Pu-Be is located at 42.5 cm above the ground level, it is demonstrated that the zone bordered by the height interval between 32 and 52 cm yields the highest EDRs.
A smoothed version of Fig.~\ref{doselateraltotal}, which is called the density of total EDR, is illustrated in Fig.~\ref{dosedensitylateraltotal}, and we see that the outlet locations of horizontal channels are emphasised in terms of the EDR intensity, while the smooth dose patterns in Fig.~\ref{dosedensitylateraltotal} substitute for the details obtained in Fig.~\ref{doselateraltotal}.
\subsection{The top surface}
\begin{figure}[H]
\begin{center}
\setlength{\belowcaptionskip}{-4ex} 
\includegraphics[width=12cm]{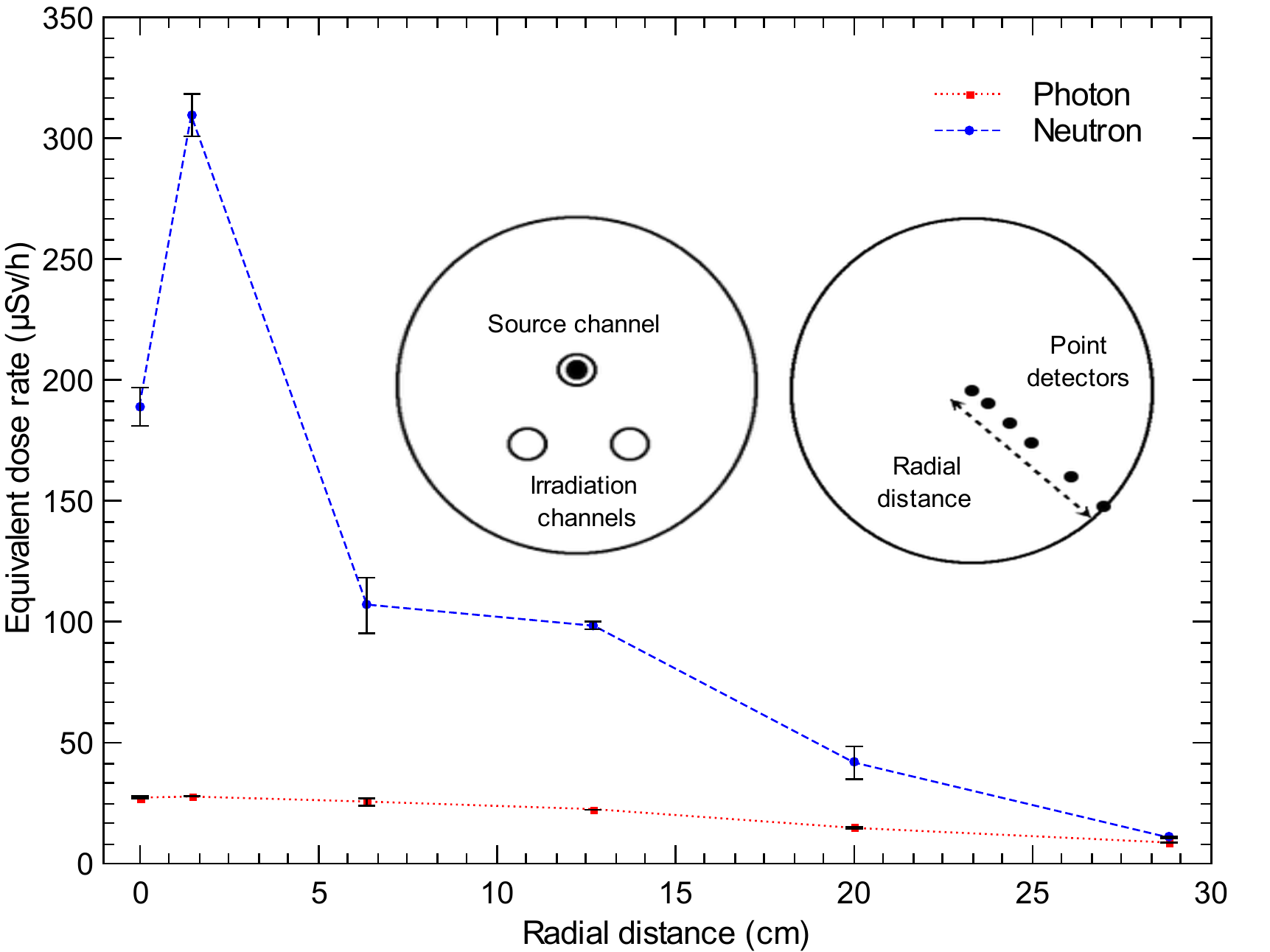}
\caption{Radial variation of EDR along the outlet of the vertical channel by using six point detectors.}
\label{Topdoseplot}
\end{center}
\end{figure}
Another surface that is described in Fig.~\ref{decomp} is the top surface, and we follow the same procedure for this surface that covers the outlets of all the vertical channels together with the cross-sectional area of the plexiglass rod in the source channel.
Recalling that the 5-Ci ${}^{239}$Pu-Be is installed at 42.5 cm and the height of the neutron howitzer is 72 cm, the top surface has a significantly higher potential for the radiation exposure with respect to the lateral surface. Especially, the small gap between the plexiglass rod and the exit of the source channel requires particular attention. The radial variation of EDR for neutrons and photons on the top surface of the NH3 howitzer is represented in Fig.~\ref{Topdoseplot}, where the radial distance is defined as the Pythagorean addition of the vertical distance and the horizontal distance introduced in Fig.~\ref{decomp}. The first arresting point in Fig.~\ref{Topdoseplot} is the sharp rise of the EDR for neutrons during the progression from the plexiglass zone to the aforementioned aperture. This area constitutes the highest equivalent dose field among all the tracked locations on both surfaces, attaining the EDRs as high as 300 \textmu Sv/h. As it is also observed on the lateral surface, the EDRs for neutrons remarkably exceed those for photons on the top surface of the NH3 howitzer except the borderlines. Particularly, the EDR for neutrons at the outlet of the vertical irradiation channel is 98.317$\pm$1.376 \textmu Sv/h, whilst it is 22.479$\pm$0.067 \textmu Sv/h for photons. 
\begin{figure}[H]
\begin{center}
\setlength{\belowcaptionskip}{-4ex} 
\includegraphics[width=10.5cm]{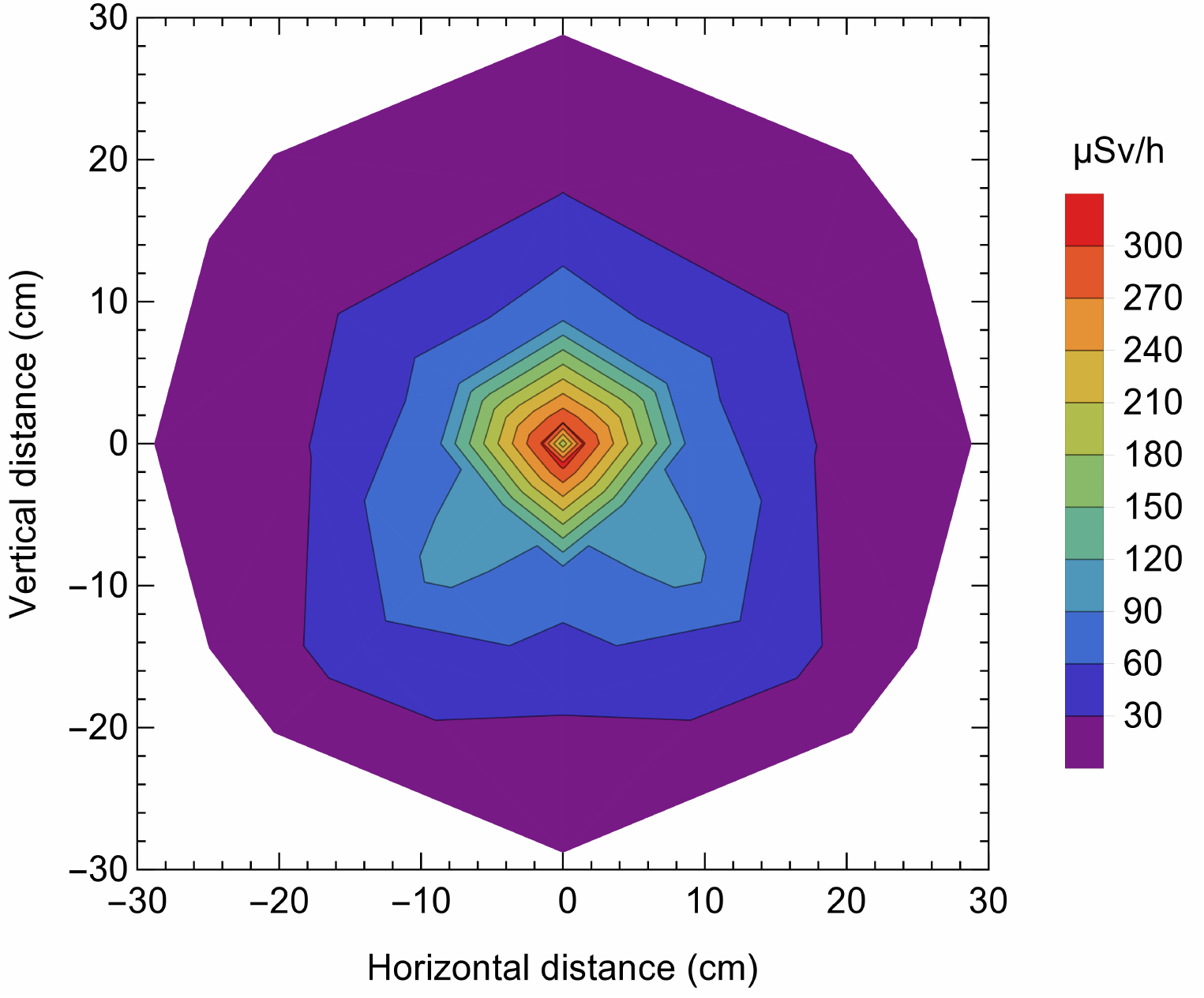}
\caption{Contours of EDR for neutrons on the top surface of the NH3 howitzer.}
\label{dosetopneut}
\end{center}
\end{figure}
Again, we plot the EDR contours and we first illustrate the EDRs for neutrons on the top surface of NH3 howitzer in Fig.~\ref{dosetopneut}. It is found that the dose patterns expand from the source channel towards the outlets of vertical irradiation channels that are clearly pointed out by the conical shapes. 
\begin{figure}[H]
\begin{center}
\setlength{\belowcaptionskip}{-4ex} 
\includegraphics[width=10.5cm]{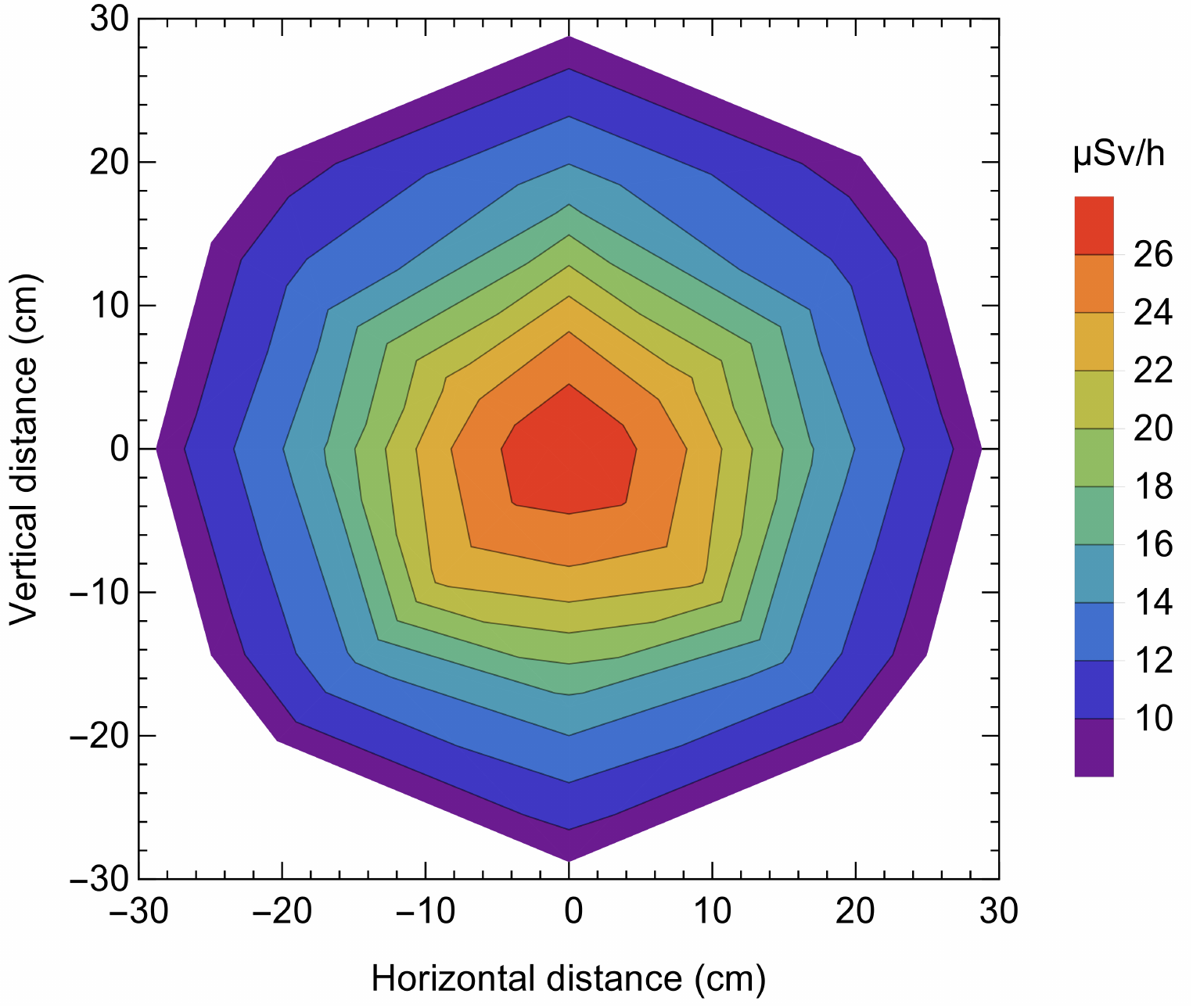}
\caption{Contours of EDR for photons on the top surface of the NH3 howitzer.}
\label{dosetopphot}
\end{center}
\end{figure}
\noindent
The dose profile in Fig.~\ref{dosetopneut} also implies that the semi-circular irradiation region, where the vertical irradiation channels are positioned, is distinctly more dangerous in comparison with the remaining part.\\
The latter contour plot for the top surface of the NH3 howitzer is dedicated to the EDRs for photons, and Fig.~\ref{dosetopphot} demonstrates that the photon EDRs on the top surface do not show a significant anisotropy despite the existence of the vertical irradiation channels. Unlike the EDRs for neutrons shown in Fig.~\ref{dosetopneut}, the EDRs for photons yield clear dose bands that monotonically decrease when moving away from the source channel.\\
\begin{figure}[H]
\begin{center}
\setlength{\belowcaptionskip}{-4ex} 
\includegraphics[width=10.5cm]{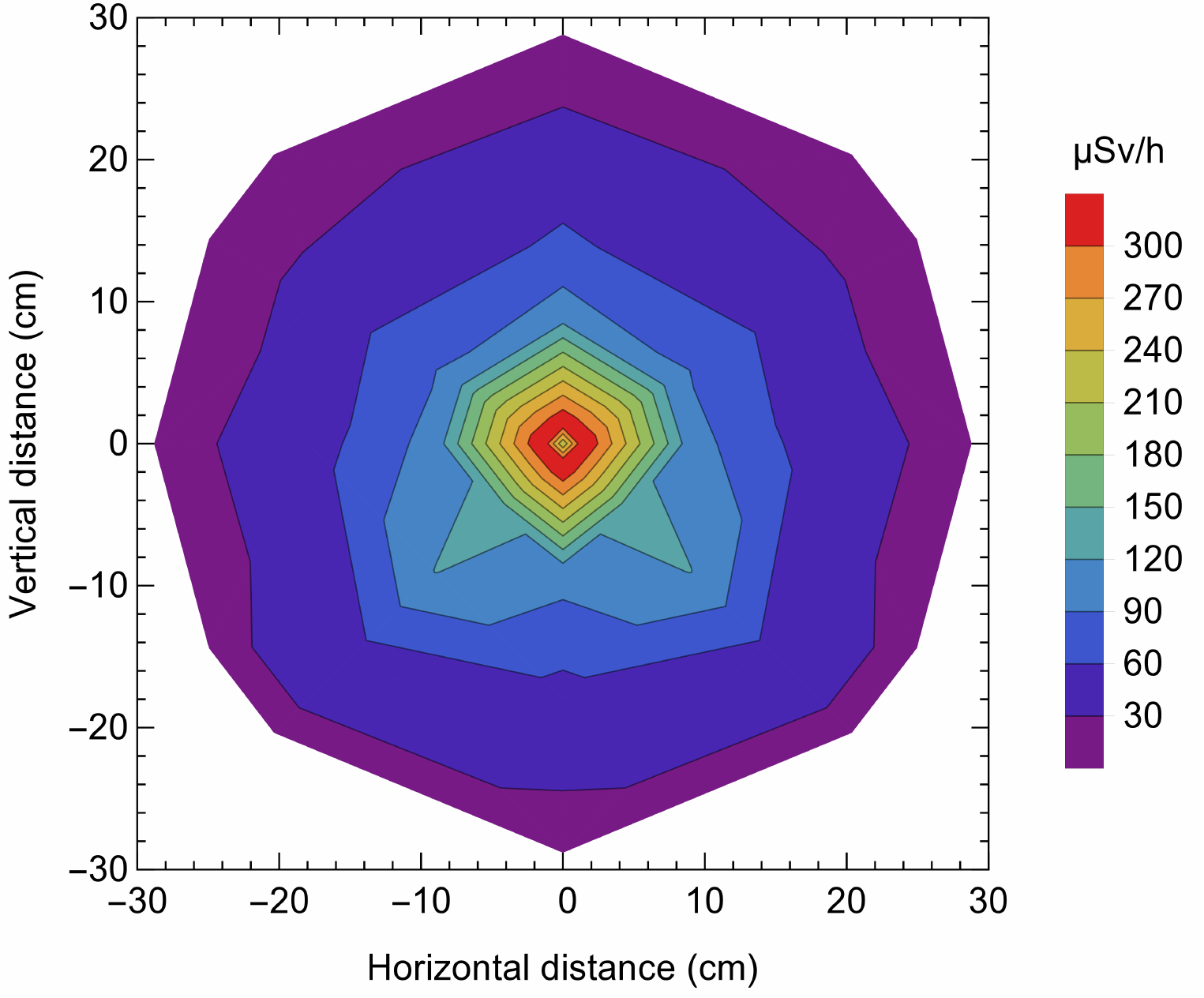}
\caption{Contours of total EDR on the top surface of the NH3 howitzer.}
\label{dosetoptotal}
\end{center}
\end{figure}
In the final contour plot, we superimpose the EDRs of photons onto the EDRs of neutrons as shown in Fig.~\ref{dosetoptotal} in order to obtain the resultant profile. As we have previously seen the similar characteristic in Fig.~\ref{doselateraltotal} for the lateral surface, the similarity between the neutron EDRs presented in Fig.~\ref{dosetopneut} and the total EDRs depicted in Fig.~\ref{dosetoptotal} proves that the resultant EDRs are mainly controlled by the neutron EDRs for the top surface.
\begin{figure}[H]
\begin{center}
\setlength{\belowcaptionskip}{-4ex} 
\includegraphics[width=11cm]{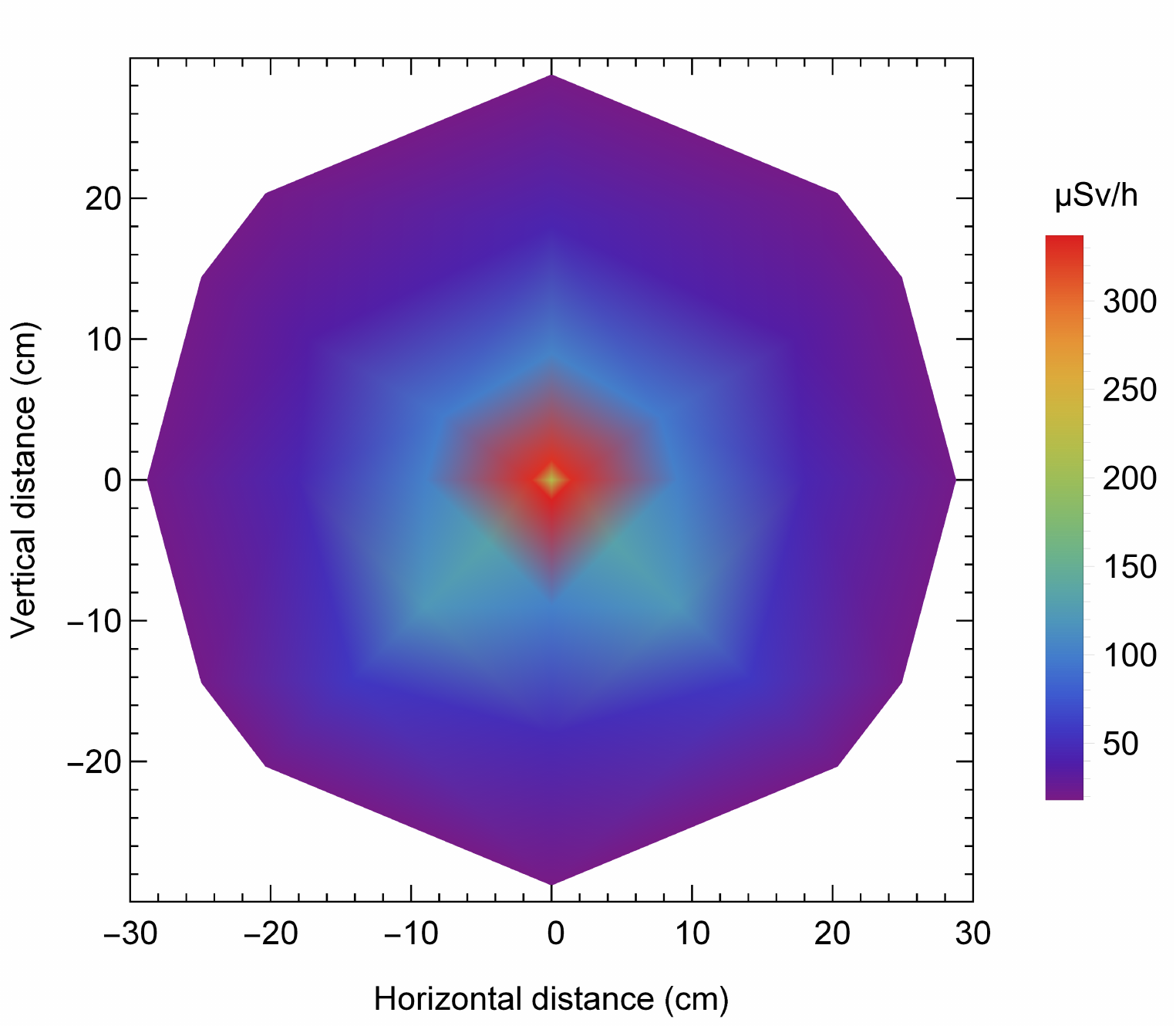}
\caption{Density of total EDR on the top surface of the NH3 howitzer.}
\label{dosedensitytopltotal}
\end{center}
\end{figure}
A smoothed representative of Fig.~\ref{dosetoptotal} is reproduced in Fig.~\ref{dosedensitytopltotal}, and we see that among the highest dose fields are the periphery of the source channel and the outlet locations of the vertical channels. 
\section{Conclusion}
Upon our Monte Carlo simulations for the NH3 howitzer, we have observed that the equivalent dose rates for photons are not negligible although both qualitative and quantitative profiles of the total equivalent dose rates are relatively governed by the equivalent dose rates of neutrons. Finally, the top surface of the NH3 howitzer is more hazardous than the lateral surface in terms of the radiation dose field especially due to the presence of the small gap between the source channel and the plexiglass rod, while both the horizontal and vertical irradiation channels deliver approximately the same equivalent dose rates.   
\bibliographystyle{elsart-num}
\nocite{*}
\bibliography{manu3Dsurfarx}
\end{document}